\documentclass[12pt]{iopart}

\usepackage{iopams}
\usepackage{amsfonts}
\usepackage{amssymb}
\usepackage{calc}
\usepackage{graphicx}
\usepackage{bm}
\usepackage{color}

\def\unit#1{\mathbf{#1}}
\def\be{\begin{equation}}
\def\ee{\end{equation}}
\def\bea{\begin{eqnarray*}}
\def\eea{\end{eqnarray*}}
\def\ba{\begin{array}}
\def\ea{\end{array}}
\def\bse{\begin{subequations}}
\def\ese{\end{subequations}}

\def\H{\mathbf{H}}

\def\P{\mathbf{P}}

\def\fs{\mathrm{fs}}

\def\chirp{\varphi_2}

\begin{document}

\title{Coherent strong-field control of multiple states by a single chirped femtosecond laser
pulse}

\author{M Krug, T Bayer, M Wollenhaupt, C Sarpe-Tudoran and T Baumert}
\address{Universit\"at Kassel, Institut f\"ur Physik und CINSaT,
Heinrich-Plett-Str. 40, D-34132 Kassel, Germany}

\eads{\mailto{wollenha@physik.uni-kassel.de},
\mailto{tbaumert@physik.uni-kassel.de}}

\author{S S Ivanov$^{1}$ and N V Vitanov$^{1,2}$}
\address{$^1$ Department of Physics, Sofia University, James
Bourchier 5 blvd, 1164 Sofia, Bulgaria}
\address{$^2$ Institute of Solid
State Physics, Bulgarian Academy of Sciences, Tsarigradsko
chauss\'{e}e 72, 1784 Sofia, Bulgaria}

\ead{vitanov@phys.uni-sofia.bg}

\begin{abstract}
We present a joint experimental and theoretical study on
strong-field photo-ionization of sodium atoms using chirped
femtosecond laser pulses. By tuning the chirp parameter, selectivity
among the population in the highly excited states $5p$, $6p$, $7p$
and $5f$, $6f$ is achieved. Different excitation pathways enabling
control are identified by simultaneous ionization and measurement of
photoelectron angular distributions employing the velocity map
imaging technique.  Free electron wave packets at an energy of
around 1~eV are observed. These photoelectrons originate from two
channels. The predominant 2+1+1 Resonance Enhanced Multi-Photon
Ionization (REMPI) proceeds via the strongly driven two-photon
transition $4s \leftarrow\leftarrow3s$, and subsequent ionization
from the states $5p$, $6p$ and $7p$ whereas the second pathway
involves 3+1 REMPI via the states $5f$ and $6f$. In addition,
electron wave packets from two-photon ionization of the non-resonant
transiently populated state $3p$ are observed close to the
ionization threshold. A mainly qualitative five-state model for the
predominant excitation channel is studied theoretically to provide
insights into the physical mechanisms at play. Our analysis shows
that by tuning the chirp parameter the dynamics is effectively
controlled by dynamic Stark-shifts and level crossings.  In
particular, we show that under the experimental conditions the
passage through an uncommon three-state ``bow-tie'' level crossing
allows the preparation of coherent superposition states.
\end{abstract}

 \maketitle

\section{Introduction}\label{introduction}

Selective excitation of preselected target states making use of
shaped femtosecond laser pulses is at the heart of coherent quantum
control
\cite{Baumert1997Wiley,Rice2000Wiley,Shapiro2003Wiley,Rabitz2000Science,Levis2002JPCA,Dantus2004ChemRev,Wollenhaupt2005ARPC,Brixner2005Springer,Tannor2006USB,Fielding2008JPB}.
Closed-loop optimization strategies
\cite{Rabitz2000Science,Judson1992PRL,Assion1998Science,Baumert1997APB,Meshulach1997OptCom,Bardeen1997CPL}
have proven enormously successful in controlling a huge variety of
quantum systems, however studies on model systems employing defined
pulse shapes are the key to better understand the underlying
physical mechanisms and to further develop quantum control concepts
and techniques. This applies in particular to \emph{strong-field}
quantum control
\cite{Wollenhaupt2003PRA,Dudovich2005PRL,Herrero2006PRL,Zhdanovich2008PRL,Bayer2009PRL}
characterized by non-perturbative interaction of a quantum system
with intense shaped laser pulses. Strong-field physical mechanisms
involve---besides the interference of multiple excitation
pathways---adiabatic and non-adiabatic time evolution accompanied by
Dynamic Stark-Shifts (DSSs) in the order of hundreds of meV. The
latter is responsible for modification of the atomic states or
molecular potential surfaces
\cite{Frohnmeyer1999CPL,Sussman2006Science,Sola2006CPL} such that
new pathways become available and new target states---inaccessible
in weak laser fields---open up. Recent studies of strong-field
control on model systems devoted to the analysis of the basic
physical mechanisms revealed that the concept of Selective
Population of Dressed States (SPODS) \cite{Wollenhaupt2006CPL}
provides a natural description of controlled dynamics in intense
shaped laser fields. For example, it was shown that ultrafast
switching among different target channels by phase discontinuities
within the pulse
\cite{Wollenhaupt2003PRA,Wollenhaupt2006CPL,Wollenhaupt2006PRA,Wollenhaupt2006JPPA},
Rapid Adiabatic Passage (RAP) by chirped pulses
\cite{Wollenhaupt2006APB} and combinations thereof
\cite{Bayer2008JPB} are realizations of this general concept.

Chirped pulses are a well-established tool in quantum control
because they usually serve as a prototype for shaped pulses with
controllable envelope and time-varying instantaneous frequency.
Therefore, they have played a prominent role in the development of
quantum control concepts and techniques and are still the
``workhorse'' to test novel strategies in quantum control. Examples
of quantum control with chirped pulses comprise studies of selective
excitation and ionization of multi-level system in alkali atoms
\cite{Zhdanovich2008PRL,Wollenhaupt2006APB,Melinger1992PRL,Balling1994PRA,Chatel2003PRA,Djotyan2003PRA,Nakajima2007PRA,Clow2008PRL},
control of molecular dynamics in diatomics and dyes
\cite{Ruhman1990JOSAB,Chelkowski1990PRL,Melinger1991JCP,Bardeen1995PRL,Assion1996CPL,Lozovoy1998CPL,Hashimoto2003APL,Nuernberger2007PCCP},
measurement of coherent transients \cite{Monmayrant2006PRL} and the
development of adiabatic passage techniques \cite{Vitanov2001ACPC}.

In the present contribution we employ chirped ultrashort laser
pulses resulting from phase modulation of the laser spectrum to
study Resonance Enhanced Multi-Photon Ionization (REMPI) of a
multi-level system in sodium atoms. We demonstrate experimentally,
that different excitation pathways and, accordingly, different
target channels can be addressed selectively by a single control
parameter, i.e. the chirp. The nature of these pathways is unraveled
by measurement of Photoelectron Angular Distributions (PADs) from
Velocity Map Imaging (VMI)
\cite{Chandler1987JChemPhys,Bordas1996RevSciInstrum,Eppink1997RevSciInstrum,Winterhalter1999JChemPhys,Vrakking2001RevSciIntrum,Whitaker2003CUP},
yielding detailed information on the origin of the released
photoelectron wave packets. Theoretical investigations of the
light-atom interaction reveal an interplay of different physical
mechanisms governing control. Analysis of the neutral excitation
dynamics for a five-state model-atom (including the most relevant
states $3s$, $4s$, $5p$, $6p$ and $7p$) under influence of a chirped
ultrashort laser pulse highlights how physical mechanisms, such as
RAP and DSS, act jointly to either address single states among the
high lying sodium states $5p$, $6p$ and $7p$ (cf.
Fig.~\ref{Fig:Exscheme}), or excite superpositions of any two
neighboring states. We point out that the present paper extends two
earlier techniques in several significant directions. The technique
of Melinger \textit{et al.} \cite{Melinger1992PRL} uses a single
chirped picosecond laser pulse to selectively excite the two
fine-structure components $3p_{1/2}$ and $3p_{3/2}$ in sodium atoms.
The present technique adds a DSS to the control tools, which enables
the population of a third state, and also the creation of coherent
superposition states. The technique of Clow \textit{et al.}
\cite{Clow2008PRL} makes use of a shaped femtosecond pulse to
selectively populate a single highly excited state. The present
technique is more flexible, since it allows to populate several
different states by variation of a single parameter: the chirp.

The article is organized as follows. We start in
Sec.~\ref{Sec:ExpLayout} by introducing the excitation and
ionization scheme of sodium atoms exposed to ultrashort
near-infrared laser pulses, and subsequently describe the details of
our experimental setup. The experimental results are presented in
Sec.~\ref{Sec:ExpResults} along with a physical discussion of
general features observed in the measured PADs supported by
numerical simulations of the measurement results. Sec.~\ref{system}
provides a detailed theoretical analysis of the strong-field induced
chirped excitation dynamics in terms of adiabatic states,
highlighting different physical mechanisms that govern the
light-atom interaction. We conclude the paper with a brief summary
and conclusions.

\section{Experiment}\label{Sec:ExpLayout}

In our experiment, we combine spectral phase shaping to produce
chirped ultrashort laser pulses with the measurement of PADs
resulting from REMPI of sodium atoms, employing the VMI technique.
In this section, we first introduce the sodium excitation scheme
with emphasis on the different accessible excitation and ionization
pathways. Then we describe the experimental setup and layout of our
photoelectron imaging spectrometer.

\subsection{Excitation scheme} \label{Sec:ExScheme}

\begin{figure}[tbfph]
\begin{center}
\includegraphics[width=10cm]{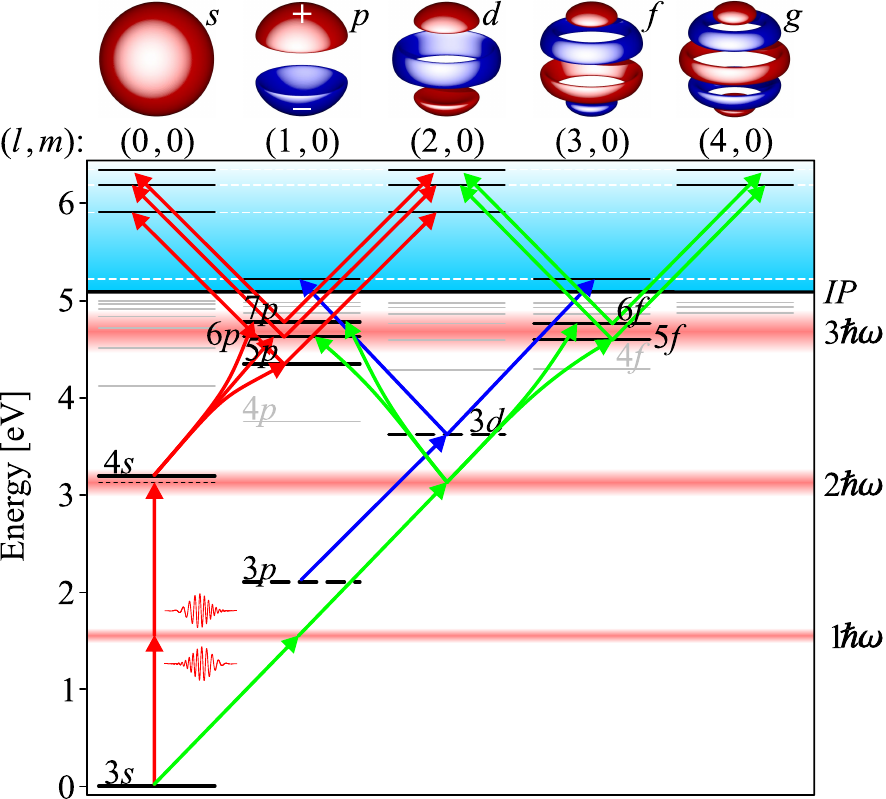}
\end{center}
\caption{\label{Fig:Exscheme} (Color online) excitation and
ionization scheme of sodium atoms illustrating the excitation
pathways that arise during the interaction with an intense 795~nm,
30~fs FWHM laser pulse. These pathways comprise a 2+1+1 REMPI (red
arrows) and a 3+1 REMPI (green arrows) process from the $3s$ ground
state as well as a two-photon ionization process from state $3p$
(blue arrows). Blurred red bars represent the one, two and
three-photon spectrum of our laser respectively. Since state $4s$
lies within the bandwidth of the two-photon spectrum, the laser
strongly drives the transition $4s\leftarrow\leftarrow3s$. Once
state $4s$ is populated, population flows to states $5p$, $6p$ and
$7p$, giving rise to photoelectron wave packets with combined $s$
and $d$-symmetry at characteristic kinetic energies 0.76~eV, 1.04~eV
and 1.20~eV in the ionization continuum. A competing excitation
pathway is opened up by three-photon absorption leading to
population of states $5f$ and $6f$ in addition. Photoelectrons from
this excitation channel are characterized by a combined $d$ and
$g$-symmetry of the measured PADs at kinetic energies 1.02~eV and
1.18~eV respectively. Two-photon ionization from the non-resonant,
transiently populated state $3p$ results in photoelectron wave
packets at about 0.2~eV, having combined $p$ and $f$-symmetry. For
illustrative purposes, the relevant symmetries of the released
photoelectron wave packets are visualized on top of the figure in
red and blue, encoding the positive and negative sign of the
electron wave function respectively.}
\end{figure}

Fig.~\ref{Fig:Exscheme} shows the excitation and ionization scheme
of sodium atoms based on energy level information taken from the
NIST-database \cite{NIST}. Different multi-photon excitation
pathways are accessible during the interaction of sodium atoms with
intense ultrashort laser pulses (laser specifications are given in
Sec.~\ref{Sec:ExpSetup}). The predominant excitation pathway is a
2+1+1 REMPI process via the two-photon transition
$4s\leftarrow\leftarrow3s$ (red arrows in Fig.~\ref{Fig:Exscheme})
which is nearly resonant with our laser spectrum
\cite{Praekelt2004PRA}. Consequential population of states $5p$,
$6p$ and $7p$ gives rise to photoelectron wave packets in the
ionization continuum having $s$ or $d$-symmetry. The recorded PADs
therefore exhibit a combined $s$ and $d$-symmetry and are measured
at the distinct kinetic energies 0.76~eV, 1.04~eV and 1.20~eV,
corresponding to states $5p$, $6p$ and $7p$ respectively.
Alternatively, a 3+1 REMPI process (green arrows in
Fig.~\ref{Fig:Exscheme}) based on three-photon absorption from the
$3s$ ground state with no intermediate resonances is taken into
account, contributing also to the population of states $5p$, $6p$
and $7p$ but, in addition, transferring population to states $5f$
and $6f$. One-photon ionization of the latter results in
photoelectron wave packets with $d$ and $g$-symmetry at kinetic
energies 1.02~eV and 1.18~eV respectively. These photoelectrons are
distinguished from the $p$ state contributions (at 1.04~eV and
1.20~eV) by the symmetry of their angular distributions. In the
following we will refer to the different photoelectron contributions
as different \emph{energy channels} at nominal kinetic energies of
0.8~eV, 1.0~eV and 1.2~eV, and infer their origin, i.e. the
excitation pathway, from the \emph{angular} distribution. Both
multi-photon excitation pathways proceed via the intermediate,
non-resonant state $3p$, which is only transiently populated.
However, since ionization takes place \emph{during} the excitation
also photoelectrons from this state are detected at low kinetic
energies around 0.2~eV (blue arrows in Fig.~\ref{Fig:Exscheme}). For
more details see caption of Fig.~\ref{Fig:Exscheme}.

\subsection{Setup}\label{Sec:ExpSetup}

\begin{figure}
\begin{center}
\includegraphics{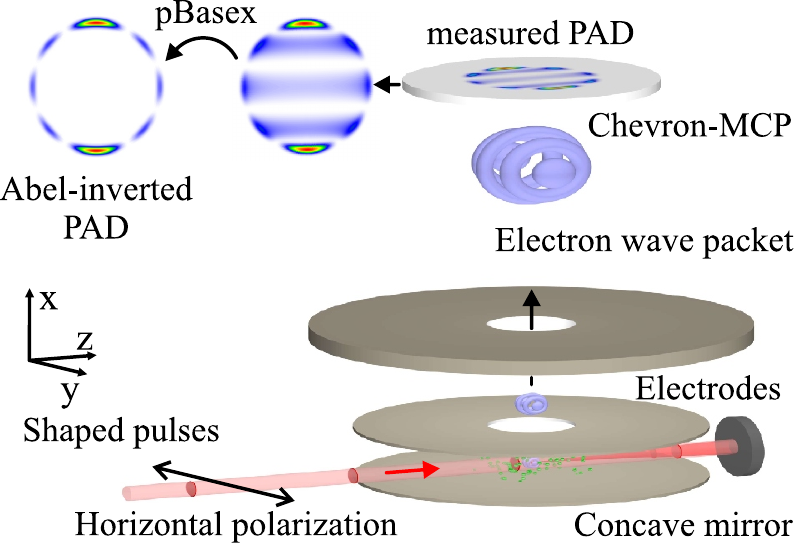}
\end{center}
\caption{\label{Fig:ExSetup} (Color online) experimental setup.
Horizontally polarized femtosecond laser pulses are sent into a
vacuum chamber and refocused by a 50~mm on-axis concave mirror into
sodium vapor provided by an alkali metal dispenser source (not
shown). Photoelectrons emitted by the light-atom interaction are
projected towards a position sensitive MCP-detector using the VMI
method. The amplified signal is recorded by a 1.4 million pixels
camera-system and sent to a computer. An Abel-inversion is performed
using the pBasex-algorithm.}
\end{figure}

In this section the experimental setup comprising the laser system
and the photoelectron imaging spectrometer is described. Intense
795~nm, 30~fs FWHM (Full Width at Half Maximum) laser pulses
provided by an amplified 1~kHz Ti:sapphire laser system
(\emph{Femtolasers Femtopower Pro}) were phase modulated in
frequency domain by a home-built pulse shaper
\cite{Praekelt2003RevSciInstrum}, applying quadratic phase masks of
the form $\varphi_{mod}(\omega)=\chirp/2\cdot(\omega-\omega_0)^2$,
where $\omega_0$ is the central frequency of our laser spectrum
\cite{Wollenhaupt2006APB}. The chirp parameter $\chirp$ was varied
in the range from $-2000~\mathrm{fs}^2$ to $+2000~\mathrm{fs}^2$ in
steps of $\Delta\chirp=100~\mathrm{fs}^2$. The chirped output pulses
of 12~$\mu$J energy were sent into a vacuum chamber and refocussed
by a concave mirror (5~cm focal length; we estimated a \emph{peak}
intensity of about $10^{13}$~W/cm$^2$ for the bandwidth-limited
pulse) into sodium vapor supplied by an alkali metal dispenser
source, as shown in Fig.~\ref{Fig:ExSetup}. Photoelectrons released
during the strong-field interaction of the shaped pulses with single
atoms were detected by a photoelectron imaging spectrometer using
the VMI method. In order to compensate the residual chirp of the
unmodulated pulse, we performed an \emph{in situ} adaptive
optimization of the multi-photon ionization of water vapor
background (about $4\times10^{-7}$~mbar) in the interaction region
of the spectrometer. The resulting optimal compensation phase was
additionally applied to the pulse shaper during the experiments,
ensuring an error in the chirp parameter $\chirp$ of less than
150~fs$^2$. The energy calibration of the imaging spectrometer was
performed using a 3+1 REMPI of xenon atoms excited by a Nd:YAG ns
laser system at 355~nm, achieving a spectrometer resolution of
60~meV at 0.5~eV. Employing the energy calibrated photoelectron
imaging spectrometer we studied angular and energy resolved
photoelectron spectra as a function of the chirp parameter $\chirp$.

\section{Experimental results and discussion}\label{Sec:ExpResults}

\begin{figure}
\begin{center}
\includegraphics[scale=1.5]{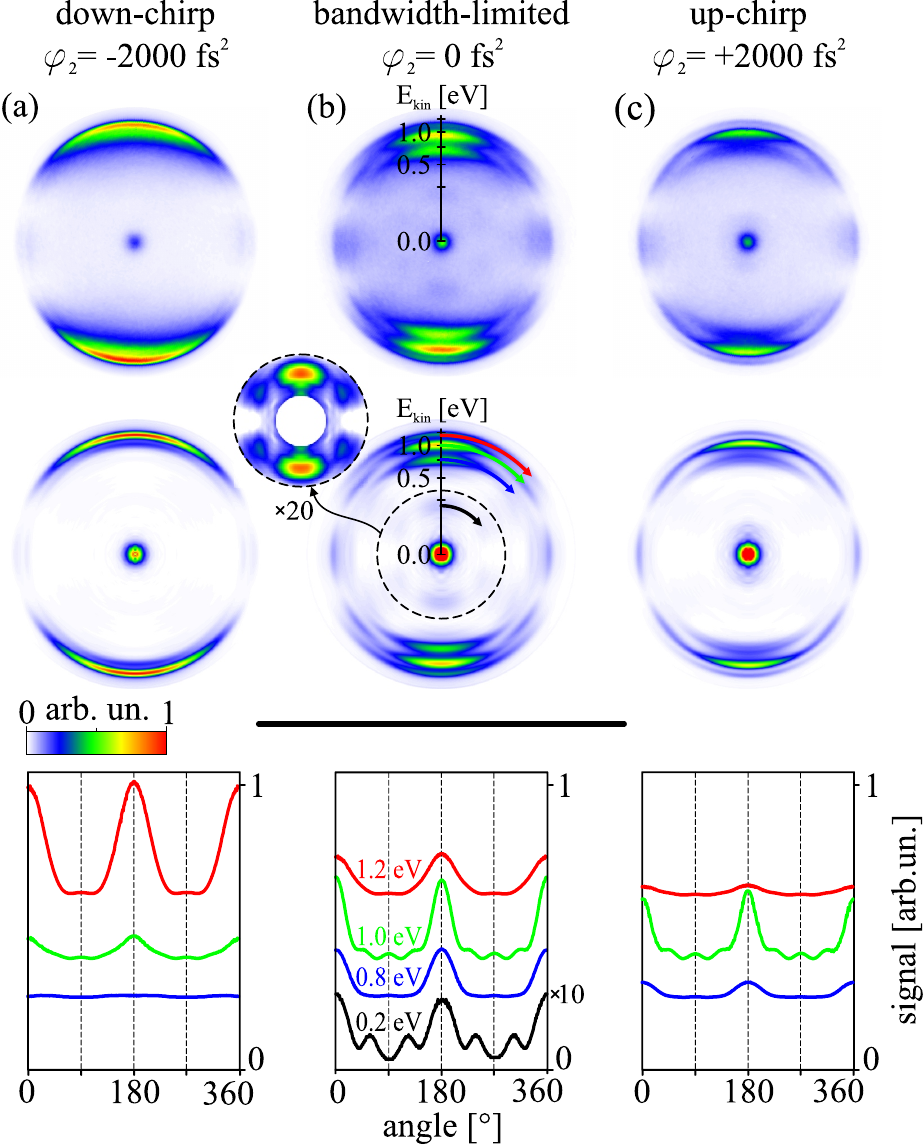}
\end{center}
\caption{\label{Fig:measure} (Color online) measured PADs from
excitation and ionization of sodium atoms using both chirped and
bandwidth-limited fs laser pulses. In the upper row measured PADs
for different values of the chirp parameter $\chirp$ are shown. (a)
$\chirp=-2000~\mathrm{fs}^2$ (down-chirp). (b) $\chirp=0$
(bandwidth-limited). (c) $\chirp=+2000~\mathrm{fs}^2$ (up-chirp).
All images are scaled to the same maximum value. The middle row
contains the corresponding Abel-inverted PADs obtained using the
pBasex-algorithm. Angular sections through the retrieved PADs at
kinetic energies of about 0.2~eV, 0.8~eV, 1.0~eV and 1.2~eV (lower
row) reveal the symmetries of the observed contributions and shed
light on the underlying ionization pathways. The signal offsets are
introduced for better visibility.}
\end{figure}

\begin{figure}
\begin{center}
\includegraphics[scale=1]{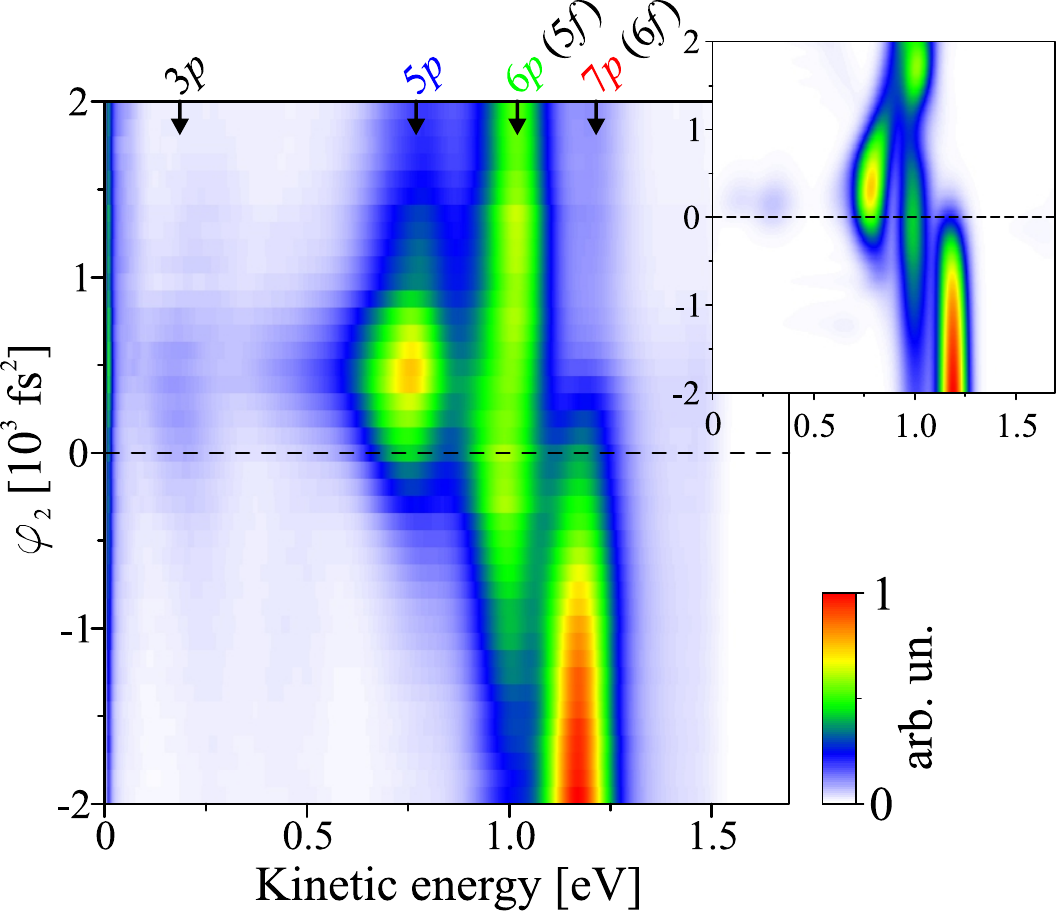}
\end{center}
\caption{\label{Fig:2dmap} (Color online) measured photoelectron
kinetic energy distributions as a function of the chirp parameter
$\chirp$. The data were obtained by angular integration of the
retrieved PADs. Three main energy channels are observed at 0.8~eV,
1.0~eV and 1.2~eV, each of which can be activated by appropriate
choice of the chirp parameter. For $\chirp\ll0$, i.e. strongly
down-chirped laser pulses, photoelectrons with high kinetic energies
related to the high lying states $7p$ (and minor $6f$ contribution)
are produced. The intermediate channel at 1.0~eV, related to states
$6p$ and $5f$, is addressed by strongly up-chirped laser pulses with
$\chirp\gg0$. Photoelectrons with kinetic energies around 0.8~eV,
corresponding to state $5p$, are favored at small positive values of
$\chirp$, i.e. high laser pulse peak intensities. The weak
contribution at 0.2~eV in the same $\chirp$-region stems from
ionization of the non-resonant state $3p$. The inset shows results
from a numerical simulation of the multi-photon excitation and
ionization process.}
\end{figure}

Figure~\ref{Fig:measure} (upper row) shows measured PADs from REMPI
of sodium atoms with chirped fs laser pulses for three exemplary
values of the chirp parameter $\chirp$. The middle row displays the
corresponding Abel-inverted (retrieved) PADs obtained by employing
the pBasex-algorithm \cite{Whitaker2003CUP,Garcia2004RevSciIntrum}.
When PADs arise from ionization with polarization shaped pulses
\cite{Wollenhaupt2009APB}, direct tomography methods have been
developed for three-dimensional reconstruction of ultrashort free
photoelectron wave packets \cite{Wollenhaupt2009APBR}. Angular
sections through the retrieved PADs at kinetic energies 0.8~eV,
1.0~eV and 1.2~eV, as plotted in the lower row, serve to identify
the symmetry of the different energy channels observed in the PADs.
The PAD measured for the unmodulated, i.e., bandwidth-limited pulse
is depicted in the central column. Three major contributions are
observed at kinetic energies 0.8~eV, 1.0~eV and 1.2~eV, related to
the energy channels discussed above (cf. Sect.~\ref{Sec:ExScheme}).
The angular section taken at 1.2~eV exhibits two minor nodes between
$0^\circ$ and $180^\circ$, i.e. $d$-symmetry. This channel is
attributed mainly to ionization via state $7p$ (red excitation
pathway in Fig.~\ref{Fig:Exscheme}), though our numerical
simulations (inset of Fig.~\ref{Fig:2dmap}) indicate, that also
ionization via state $6f$ (green excitation pathway in
Fig.~\ref{Fig:Exscheme}) delivers a minor contribution. The
contribution of an $s$-wave to this channel, as expected from the
excitation scheme Fig.~\ref{Fig:Exscheme}, is reflected in the
\emph{weak} equatorial signal: At an angle of $90^\circ$ $s$ and
$d$-wave have opposite sign and, thus, interfere destructively,
whereas at the poles, i.e. at $0^\circ$ and $180^\circ$, both waves
add up constructively. The section taken at 1.0~eV exhibits 4 nodes
between $0^\circ$ and $180^\circ$, corresponding to $g$-symmetry.
This contribution originates predominantly from ionization via state
$5f$. The observation that the lobe at $90^\circ$ (and $270^\circ$
respectively) is slightly lowered with respect to its two neighbors
indicates a weak $d$-wave contribution interfering destructively
with the $g$-wave in this angular segment. The contribution measured
at 0.8~eV shows again combined $s$ and $d$-symmetry and is ascribed
to ionization via state $5p$.

Moreover, a weak contribution is observed at about 0.2~eV, a
magnification of which is shown in the inset of
Fig.~\ref{Fig:measure}(b). The nodal structure of this signal
exhibits distinct $f$-symmetry. However, the pronounced poles of the
PAD as well as the fact, that the nodes at $45^\circ$ and
$135^\circ$ in the angular section are raised with respect to the
node at $90^\circ$ give a hint on a $p$-wave contribution to the
photoelectron signal. Observation of photoelectron wave packets with
combined $p$ and $f$-symmetry close to the ionization threshold is
consistent with two-photon ionization from state $3p$ (blue pathway
in Fig.~\ref{Fig:Exscheme}). Note, that state $3p$ is---although
non-resonant---transiently populated during the interaction,
mediating the multi-photon processes to the state $4s$ and the high
lying $f$ states.

For large negative values of $\chirp$ (left column in
Fig.~\ref{Fig:measure}), i.e. strongly \emph{down}-chirped laser
pulses, the outer channel at kinetic energy 1.2~eV is considerably
enhanced in comparison to the bandwidth-limited case, whereas the
intermediate channel at 1.0~eV is strongly reduced and the two
innermost contributions have essentially vanished. Note the change
in symmetry of the intermediate channel which exhibits combined $s$
and $d$-symmetry in this case, indicating more efficient ionization
from state $6p$, while the $5f$ contribution is very small. Changing
the sign of $\chirp$, i.e. using strongly \emph{up}-chirped laser
pulses (right column in Fig.~\ref{Fig:measure}), suppresses the high
energy channel in favor of the intermediate channel at 1.0~eV which
dominates the PAD in this case. From its angular section at 1.0~eV
we find a combined $d$ and $g$-symmetry, as in the bandwidth-limited
case. This contribution is therefore traced back mainly to state
$5f$. The finding that the symmetry of photoelectrons from the
intermediate channel alters from $d$ to $g$ is rationalized by the
change of the ordering of red and blue frequency components within
the chirped pulse. For a down-chirped pulse, i.e. when the blue
components arrive first, initially, the system is in resonance with
the two-photon transition $4s \leftarrow\leftarrow 3s$ implying
efficient ionization via the $p$ states (red pathway in
Fig.~\ref{Fig:Exscheme}). On the other hand, up-chirped pulses favor
ionization via state $5f$ since at early times the system is in
resonance with the three-photon transition $5f
\leftarrow\leftarrow\leftarrow  3s$ (green pathway in
Fig.~\ref{Fig:Exscheme}). Such processes have also been observed in
\cite{Assion1996CPL} under different excitation conditions.

In order to provide the full picture of the chirp dependent
population flow to the different energy channels, we performed an
angular integration of all 41 measured PADs and present the
resulting energy-resolved photoelectron spectra in terms of a
two-dimensional map as a function of the kinetic energy and the
chirp parameter $\chirp$. The result obtained upon variation of
$\chirp$ in the range from $-2000~\mathrm{fs}^{2}$ to
$+2000~\mathrm{fs}^{2}$ is displayed in Fig.~\ref{Fig:2dmap}. The
three major channels at 0.8~eV, 1.0~eV and 1.2~eV are clearly
visible. Note, that for e.g. rare gas atoms under our experimental
conditions ponderomotive shifts of more than 0.5~eV are calculated.
No such shifts are observed in the experiment, since the
high-frequency approximation \cite{Avan1976JdP,LHuillier1989JOSAB}
(necessary condition for the application of the ponderomotive energy
concept) is not valid for alkalis excited by near infrared laser
radiation. An analysis of the neutral excitation dynamics behind the
observed contributions will be given in Sec.~\ref{system}. The map
illustrates the above statements, that for large negative values of
$\chirp$ the high energy channel at 1.2~eV is addressed with high
efficiency, i.e. a down-chirped pulse steers the population
predominantly towards the high lying state $7p$. For large positive
chirp values the intermediate channel is selectively addressed,
corresponding to predominant population of states $6p$ and $5f$. The
low energy channel is accessed most efficiently in the vicinity of
$\chirp=500~\mathrm{fs}^{2}$. In fact, in the regime
$0\leq\chirp\leq1000~\mathrm{fs}^2$ the photoelectron spectrum is
made up of contributions from states $5p$, $6p$ and $5f$. Because
the excitation (and simultaneous ionization) takes place on an
ultrashort time scale precluding decoherence processes, a
\emph{coherent} superposition of states $5p$, $6p$ and $5f$ is
excited in this chirp regime. Upon changing the sign of $\chirp$,
i.e. for $-1000~\mathrm{fs}^2\leq\chirp\leq0$, the laser pulse
induces a coherent superposition of states $6p$, $5f$ and $7p$.
Photoelectrons observed at about 0.2~eV for moderate positive chirps
are attributed to two-photon ionization from state $3p$.

The inset to Fig.~\ref{Fig:2dmap} shows results from a numerical
simulation of the simultaneous multi-photon excitation and
ionization process. The calculations are based on numerical
integration of the time-dependent Schr\"odinger-equation for a
neutral 20-state system (comprising those states labeled in
Fig.~\ref{Fig:Exscheme} and taking the fine structure splitting into
account) interacting with an intense chirped 795~nm, 30~fs FWHM
Gaussian input pulse. One-photon ionization from the high lying $p$
and $f$ states is treated within a simplified model employing first
order perturbation theory. We assume a flat continuum and unit
coupling elements with no additional phases for all bound-free
transitions. A more rigorous treatment of the ionization step
involving the determination of radial coupling matrix elements also
for the bound-free transitions is provided by, e.g., single-channel
quantum defect theory \cite{Seaton1983RPP} as reported for instance
in \cite{Dixit1983PRA,Leuchs1986PRL}. In order to model the
two-photon ionization from state $3p$ proceeding, for example, via
state $3d$ as indicated by the blue pathway in
Fig.~\ref{Fig:Exscheme}, we employed second order perturbation
theory. For a more detailed description of our method see
\cite{Wollenhaupt2006PRA,Wollenhaupt2006APB,Wollenhaupt2007JPhysConfSer}.
The simulation of photoelectron spectra reproduces the main features
of the experimental results very well. This allows us to look into
the underlying \emph{neutral} excitation dynamics and follow the
population flow within the bound atomic system. We find that for
\emph{large negative chirp} $\chirp$ state $7p$ is addressed almost
selectively, while for large positive $\chirp$ values both states
$6p$ and $5f$ are populated efficiently in equal measure. The latter
is in accordance with the experimental observation of the PAD with
pronounced $g$-symmetry in the intermediate channel at 1.0~eV for
\emph{large positive chirp} (see Fig.~\ref{Fig:measure}(c)). The
most efficient excitation of state $5p$ occurs for \emph{moderate
positive chirp}. However, in this chirp regime states $6p$ and $5f$
receive comparable population confirming the observation of a PAD
with a contribution of $g$-symmetry at 1.0~eV and \emph{zero chirp}.
At \emph{moderate negative chirp}, we obtain a coherent
superposition of states $6p$, $5f$ and $7p$. Note that the weak
contribution around 0.2~eV and small positive values of $\chirp$
observed in the experiment (shown in the inset to
Fig.~\ref{Fig:measure}(b)) is also reproduced in the simulation.
Within the framework of our simulation, these photoelectrons are
ascribed to two-photon ionization from state $3p$ which receives
non-perturbative \emph{transient} population. We note, that in a
perturbative regime, ionization from this transiently populated
state could be interpreted as a transition from a \emph{virtual}
state.

In the next section, we will further investigate the neutral
population dynamics by means of a reduced atomic system in order to
rationalize the general features observed in the experiment in terms
of physical mechanisms governing the excitation process.

\section{Theoretical model}\label{system}

In this section we provide a mainly qualitative description of the
system at hand. To this end, we assume that the photoelectron signal
arises most significantly through the 2+1+1 REMPI channel (red
pathway in Fig.~\ref{Fig:Exscheme}), involving the five states $3s$,
$4s$, $5p$, $6p$ and $7p$. The idea of this reduction is to
demonstrate the basic principles influencing the dynamics of the
whole system, which become more transparent in this simplified
model, involving the most significant states for our experiment. In
this approach, we adiabatically eliminated state $3p$
\cite{Vitanov2001ACPC,Shore1990Wiley,Shore2008ActPhysSlov} because
it is off resonance and receives smaller transient population than
the other coupled states. Its presence, though, affects the
population dynamics significantly for it induces strong dynamic
Stark-shifts in the energies of states $3s$ and $4s$, which
substantially modify the energy diagram.

The quantum dynamics of this five-state system obeys the
time-dependent Schr\"{o}dinger equation
\be\label{Shr} i\hbar\frac{d}{dt}\unit{c}(t)=\H(t)\unit{c}(t). \ee
The Hamiltonian $\H(t)$ in the rotating-wave approximation, rotating
with the instantaneous laser frequency $\omega(t)=\omega_0 + 2 a t$
(see Eq.~(\ref{Eqn:Chirp}) in the appendix), is given by
\cite{Vitanov2001ACPC,Shore2008ActPhysSlov}:
\be\label{ham} \H(t)=\hbar\left[ \ba{ccccc}
\Delta_{1}-S_1     & \frac{1}{2}\Omega_{12}      & 0                   & 0                  & 0 \\
\frac{1}{2}\Omega_{12}   & \Delta_{2}-S_2        & \frac{1}{2}\Omega_{23}    & \frac{1}{2}\Omega_{24}   & \frac{1}{2}\Omega_{25}\\
0                  & \frac{1}{2}\Omega_{23}      & \Delta_{3}          & 0                  & 0\\
0                  & \frac{1}{2}\Omega_{24}      & 0                   & \Delta_{4}         & 0\\
0                  & \frac{1}{2}\Omega_{25}      & 0                   & 0                  & \Delta_{5}\\
\ea\right].\ee
Here the explicit time dependence is dropped for ease of notation.
The vector $\unit{c}(t)=\left[c_1(t),c_2(t),\ldots,c_5(t)\right]^T$
consists of the amplitudes of the five states, ordered as shown
above, which are obtained by numerical integration of the
Schr\"{o}dinger-equation (\ref{Shr}), the respective populations are
$P_n(t)=\left|c_n(t)\right|^2$, $\Delta_n(t)=\omega_n-k\,\omega(t)$
are the generally time-dependent atom-laser detunings, where
$\omega_n$ are the atomic state eigenfrequencies, with $\omega_{3s}$
taken as zero, $k$ is the transition order,
$\Omega_{2n}=d_{2n}\Omega_0f(t)$ represent the one-photon couplings
of state $2$ to state $n$ ($n=3,4,5$), $\Omega_{12}=q_{12}\Omega_0^2
f^2(t)$ is the two-photon coupling between states 1 and 2, with
$f(t)$ being the chirped laser electric field envelope, $d_{mn}$ are
the relevant transition dipole moments in atomic units, $q_{12}$ is
the effective two-photon transition moment (cf. Eq.~(\ref{Q12})) and
$S_1$ and $S_2$ represent the DSS of states $1$ and $2$,
respectively, \be S_1=\frac{\Omega_{3s3p}^2}{4\Delta_{3p}},\quad
S_2=\frac{\Omega_{3p4s}^2}{4\Delta_{3p}}. \ee
The effect of the DSS due to state $3d$ is neglected for it is very
weakly coupled to the states whose energies it might influence: the
$p$ states are coupled about 10 times stronger to state $4s$ as
compared to state $3d$; state $3d$ is not directly coupled to state
$3s$, but rather through a two-photon transition. In the first two
diagonal elements of the Hamiltonian (associated with the energies
of states $3s$ and $4s$) the atom-laser detuning and the DSS add up
to a time-dependent \emph{effective} chirp: the former resulting
from the time-dependent instantaneous laser frequency $\omega(t)$,
and the latter deriving from the time-dependent shift of the level
energies due to DSS.

\begin{figure*}[tbh]
\centering
\includegraphics[angle=0,width=15.5cm]{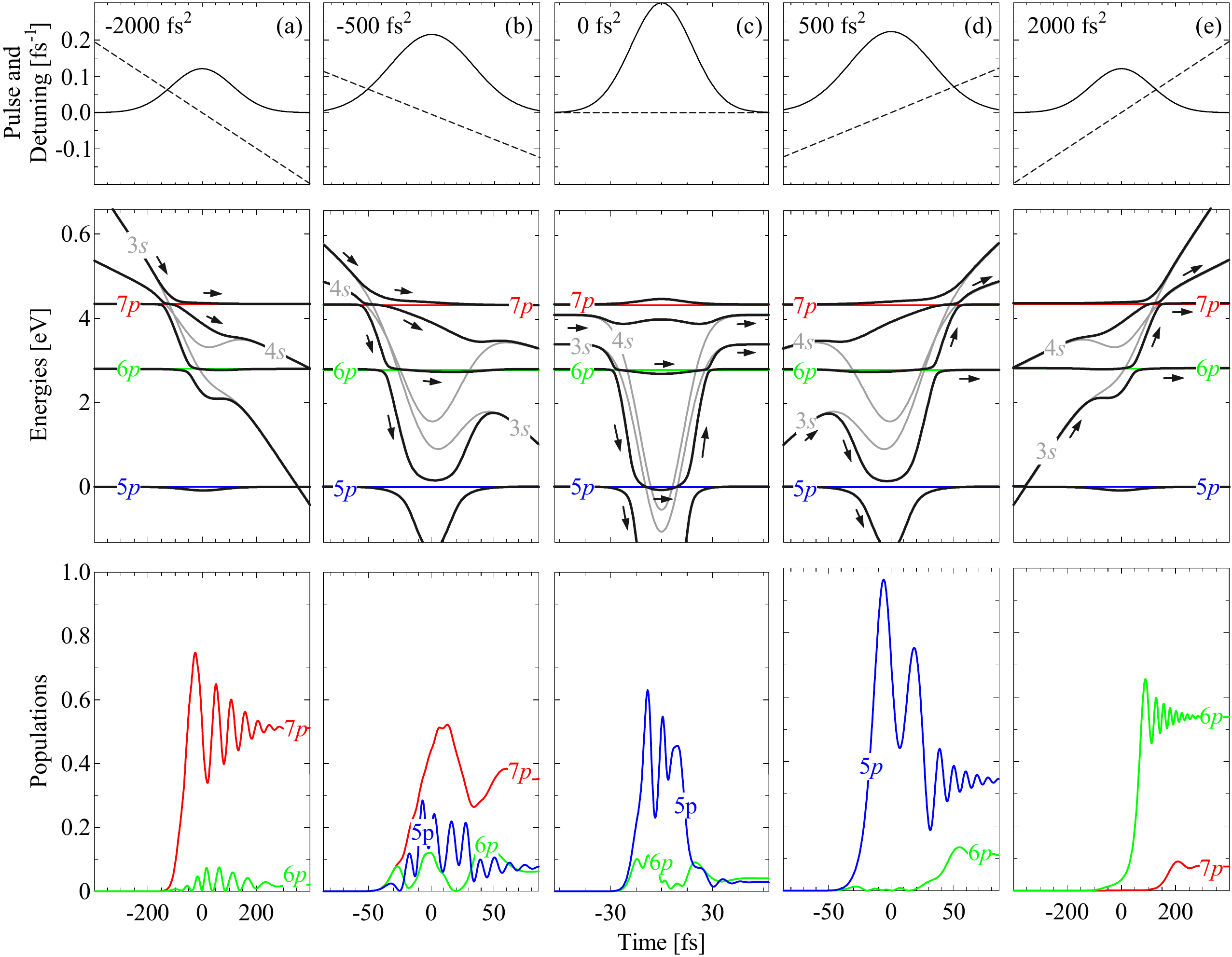}
\caption{(Color online) populations (lower frames) and energies
(middle frames) of the states of interest $5p$, $6p$ and $7p$ vs
time for $\chirp$ varied (from left to right) between $-2000~\fs^2$
(down-chirp) and $2000~\fs^2$ (up-chirp), $\Omega_0=0.3~\fs^{-1}$
and $\Delta t=30~\fs$. In the middle frames, colored and grey lines
depict the bare state energies. The latter are related to states
$3s$ and $4s$ and include the effective chirp, i.e. the chirp of the
laser as well as the chirp due to ac Stark-shifts. Black lines
represent the dressed state energies and the arrows are to show the
population flow. The populations in the lower frames are consistent
with the asymmetry in the experimental results presented in Fig.
\ref{Fig:2dmap}: for large chirps states $6p$ (positive chirp) and
$7p$ (negative chirp) are predominantly populated, whereas around
zero chirp the contribution comes mostly from state $5p$. The
envelopes (straight lines) and detunings (dashed lines) of the
modulated pulses are shown in the uppermost frames. Note that the
energies are mirrored when changing the sign of the chirp $\chirp$.}
\label{all}
\end{figure*}

\subsection{Excitation regimes}

In Fig.~\ref{all} we distinguish five different regimes in regard to
the value of the chirp $\chirp$. In all cases we plot the bare-state
energies and analyze the dynamics by accounting for the presence of
level crossings. Because it is the ionization signal that is
observed in the experiment it is also important \textit{when} a
particular level crossing occurs: a level crossing at \textit{early}
time, and the ensuing adiabatic passage transition, would translate
into a larger ionization signal than a \textit{late} crossing, where
even a significant population transfer to a certain discrete state
would not be reflected in the ionization signal.

Below we examine the dynamics of our system with particular interest
in states $5p$, $6p$ and $7p$. In Fig. \ref{all} we show the
populations and the energies of the five bare states for the chirp
$\chirp$ varied between $-2000~\fs^2$ and $2000~\fs^2$ (from left to
right) with the system initiated in state $3s$. For illustrative
purposes we pick $\Omega_0=0.3~\fs^{-1}$, corresponding to an
intensity of $3.7\times10^{12}$~W/cm$^2$ \cite{Shore1990Wiley}, and
$\Delta t=30~\fs$.

\subsubsection{Large negative chirp}
For large negative chirp ($\chirp=-2000~\fs^2$, Fig.~\ref{all}(a))
the laser field reaches resonances relative to the $7p\leftarrow4s$
(one-photon) transition and the $4s\leftarrow\leftarrow3s$
(two-photon) transition in nearly the same instant, thus creating a
``bow-tie'' level crossing pattern
\cite{Vitanov2001ACPC,Carrol1986JPA1,Carrol1986JPA2,Ostrovsky1997JPA,Harmin1991PRA,Brundobler1993JPA}
which is of particular significance because it involves three rather
than two states. This crossing results in efficient population
transfer to states $4s$ and $7p$ and depopulation of state $3s$.
Because state $7p$ is populated at such early times, it is exposed
to ionization for most of the interaction dynamics and hence has a
dominant contribution in the photoelectron signal (see
Fig.~\ref{Fig:2dmap} at 1.2~eV and $-2000$~fs$^2$).

Later on we observe almost adiabatic evolution and the population is
shared mainly between states $4s$ and $7p$ in the form of
Rabi-oscillations with fading amplitude \cite{Zamith2001PRL}. State
$6p$ acquires only marginal population mainly due to its crossing
with state $3s$ (which is, however, already depleted due to the
preceding ``bow-tie'' crossing) via a three-photon excitation
through state $4s$. The late crossings between states $3s$ and $5p$,
and also between states $4s$ and $6p$ are of no importance because
they occur after the pulse intensity has essentially vanished. State
$5p$ remains unpopulated since it is far off-resonant throughout the
entire dynamics.

\subsubsection{Large positive chirp}
For large positive chirps ($\chirp=2000~\fs^2$, Fig.~\ref{all}(e))
the energy diagram is mirrored compared to the one for large
negative chirps $\chirp$ (Fig.~\ref{all}(a)). Then initially the
system evolves adiabatically, with minor (off-resonant) population
transfer from state $3s$ to state $4s$ due to their strong mutual
coupling. Around the time of the peak laser intensity, as state $3s$
sweeps across resonance with $6p$, the latter starts to effectively
populate through the three-photon $3s-6p$ crossing. Because this
crossing occurs approximately in the middle of the laser pulse the
population of state $6p$ is exposed to ionization for a considerable
time interval, which results in significant photoelectron signal
from $6p$ (see Fig.~\ref{Fig:2dmap} at 1.0~eV and $+2000$~fs$^2$).
For the same reason---the $3s-6p$ crossing occurring near the laser
pulse maximum---the population transfer from state $3s$ to state
$6p$ is relatively efficient and only about half of the population
is left in states $3s$ and $4s$ thereafter; then only a part of this
already reduced population is transferred to state $7p$ at the
subsequent ``bow-tie'' crossing $3s-4s-7p$. Moreover, this crossing
occurs at late times and hence state $7p$ is not visible in the
photoelectron spectrum. State $5p$ remains unpopulated once again as
it stays far off any resonance.

We now turn our attention to the regimes of a moderately large chirp
$\chirp$, where the photoelectron spectrum changes from a
single-state feature to one displaying double features.

\subsubsection{Moderate negative chirp}
For a moderate negative chirp ($\chirp=-500~\fs^2$,
Fig.~\ref{all}(b)) an early crossing occurs between states $3s$,
$4s$ and $7p$ in the rising edge of the pulse, which leads to a
partial population transfer from state $3s$ to states $4s$ and $7p$,
because the laser intensity is not strong enough to enforce
adiabatic evolution. The population in state $7p$ is exposed to
ionization for the rest of the pulse, whereas the population in
state $4s$ proceeds until the subsequent $4s-6p$ crossing where it
is partially transferred to state $6p$. The leftover temporary flows
into state $5p$, which starts to emerge in the photoelectron
spectrum, and is finally driven back into state $3s$. In result, all
states $5p$, $6p$ and $7p$ are visible in the photoelectron signal,
which is an indication for the creation of a coherent superposition
of these (see Fig.~\ref{Fig:2dmap} at about $-500$~fs$^2$).

\subsubsection{Moderate positive chirp}
For moderate positive chirps ($\chirp=500~\fs^2$, Fig.~\ref{all}(d))
state $3s$ first comes very close to state $5p$ at times of the
laser pulse maximum; during this proximity the population undergoes
Rabi-type oscillations between states $3s$ and $5p$ and is exposed
to ionization from state $5p$. The signature of state $5p$ is
clearly visible and indeed, this is the regime where this state
indisputably dominates in the photoelectron signal (see
Fig.~\ref{Fig:2dmap} at 0.8~eV and $+500$~fs$^2$). In other words,
it is the DSS induced by the two-photon transition
$4s\leftarrow\leftarrow3s$, which makes the population of the
far-off-resonant state $5p$ possible \cite{Rickes2000JCP}. If this
Stark-shift were absent (e.g. if the two-photon transition
$4s\leftarrow\leftarrow3s$ were instead a single-photon one in a
gedanken scenario) state $5p$ would never receive sizeable
population. As we proceed beyond the pulse maximum state $3s$
crosses state $6p$ and the population is partially transferred to
the latter. Hence state $6p$ emerges in the photoelectron signal due
to the ensuing ionization, whereas state $7p$ is invisible in this
regime because all population left flows into state $4s$.

\subsubsection{Zero chirp.}
In this regime the laser pulse is unchirped, $\chirp=0$. Therefore,
the \emph{effective} chirp is entirely due to ac Stark-shift. The
latter is symmetric to the pulse because it is induced by the same
pulse. Moreover, because state $3s$ crosses states $6p$ and $5p$
(Fig.~\ref{all}(c)), sizeable population will visit these two states
through the respective first crossings $3s-5p$ and $3s-6p$. A second
pair of crossings in the falling edge of the pulse will induce
additional transitions $5p\leftarrow\leftarrow\leftarrow3s$ and
$6p\leftarrow\leftarrow\leftarrow3s$. The implication is that states
$5p$ and $6p$ will contribute significantly to the photoelectron
signal (see Fig.~\ref{Fig:2dmap} around $\chirp=0$). State $7p$, on
the other hand, remains well off resonance throughout and receives
only a small population due to (weak) non-resonant interaction. Its
contribution to the photoelectron signal should be therefore more
muted than these from states $5p$ and $6p$.

\subsection{Discussion}
Below we discuss the five excitation regimes in the dressed state
(adiabatic) context. When adiabatic, which demands large couplings
and low chirp rates for the avoided crossings in question, starting
in state $3s$ we end up in state $7p$ for $\chirp<0$ or in state
$6p$ for $\chirp>0$ (Fig. \ref{all}, middle frames; in the latter
case a fully non-adiabatic passage across state $5p$ occurs, since
the pulse intensity is negligible for the $3s-5p$ resonance).
Therefore, clearly from Fig. \ref{all}, our system exhibits a
somewhat adiabatic behavior for chirp $\chirp$ away from the origin.
As we get closer, the crossings shift towards the pulse wings,
whereas the pulse gets narrower in time, which in combination
results in breaking adiabaticity. The latter is further hindered by
the increased DSS, which effectively enhances the chirp rate.

We expect adiabaticity to remain almost unaffected for large
negative values of the chirp $\chirp$, since the chirp rate
$a\propto 1/\chirp$ and $\Omega\propto 1/\sqrt{\chirp}$, and to
break down for large positive values, for it relies on the
three-photon transition $6p\leftarrow\leftarrow\leftarrow3s$, which
gets weaker, as the resonances relative to $3s-4s$ and $4s-6p$
further separate in time. Larger peak intensities $\Omega_0$
strengthen adiabaticity for the transition
$7p\leftarrow\leftarrow\leftarrow3s$ and make complete population
transfer possible, as also indicated in \cite{Clow2008PRL}, whereas
for the transition $6p\leftarrow\leftarrow\leftarrow3s$ due to the
unfavorable influence of the increased DSS we predict the contrary.

\section{Summary and Conclusion}

In this contribution we presented a joint experimental and
theoretical study on strong-field Resonance Enhanced Multi-Photon
Ionization (REMPI) of sodium atoms using chirped femtosecond laser
pulses. Experimentally, Photoelectron Angular Distributions (PADs)
have proven the essential tool to identify the different excitation
and ionization pathways.

We observed three distinct ionization pathways contributing to the
measured PADs. The predominant contribution with combined $s$ and
$d$-symmetry is due to a 2+1+1 REMPI processes involving the
strongly driven two-photon transition $4s \leftarrow\leftarrow 3s$,
and subsequent ionization from the states $5p$, $6p$ and $7p$.
Photoelectrons with combined $d$ and $g$-symmetry originated from
3+1 REMPI via states $5f$ and $6f$. A weak contribution with
combined $p$ and $f$-symmetry close to the ionization threshold is
attributed to the third channel, that is two-photon ionization of
the non-resonant transiently populated state $3p$.

Selective population of the highly excited states $5p$, $6p$, $7p$
and $5f$, $6f$ was achieved by controlling a single pulse parameter,
i.e. the chirp parameter $\chirp$. In particular, we observed highly
selective population of state $7p$ using strongly down-chirped laser
pulses. For strongly up-chirped laser pulses states $6p$ and $5f$
were populated with high efficiency and a dominant signal from state
$5p$ was obtained for moderately up-chirped laser pulses. Moreover,
in the intermediate chirp regions coherent superpositions of
neighboring states have been excited.

Simulations based on numerical integration of the time-dependent
Schr\"odinger-equation for a neutral 20-state system are in
agreement with our experimental findings. In addition, a five-state
model was developed in order to provide insights into the physical
mechanisms at play. Our analysis of the time-dependent populations
showed that by tuning the chirp parameter distinct physical
mechanisms have been addressed, involving adiabatic and
non-adiabatic time evolution along with Dynamic Stark-Shifts (DSSs)
and (multiple) level crossings. It was pointed out that the
occurrence of an uncommon ``bow-tie'' level crossing is responsible
for the excitation of coherent superposition states as observed in
the experiment. The strong DSS of the two-photon transition $4s
\leftarrow\leftarrow 3s$ turned out to be of particular significance
for populating state $5p$ being inaccessible in weak laser fields.

Our results highlight the importance of studying model systems
experimentally and theoretically to better understand the physical
mechanisms of strong-field coherent control. Our findings
demonstrate that, in general, in strong-field control multiple
pathways involving different physical mechanisms are at play
simultaneously.

\appendix

\section{Details of calculations}\label{appendix}

Each $p$ state consists of $p_{1/2}$ and $p_{3/2}$ substates,
coupled by $\Omega_{1/2}$ and $\Omega_{3/2}$, respectively, to a
relevant $s$ state. Therefore initially our system comprises overall
10 states (prior to eliminating state $3p$). To simplify our
approach we perform a transformation to a dark-bright basis for each
of the $p$ states and thus eliminate half of the $p$ substates as
dark (uncoupled) states, and keep the rest, which become coupled by
the root mean square of the relevant $\Omega_{1/2}$ and
$\Omega_{3/2}$ and are the ones to be referred to as $p$ states
throughout the theoretical part of the paper.

The effective two-photon transition moment between states $3s$ and
$4s$ is
\be\label{Q12} q_{12}=-\frac{d_a d_b+d_c d_d}{2\Delta_{3p}}, \ee
where $d_{a,c}$ and $d_{b,d}$ are the dipole moments for the
transitions $3p_{1/2,3/2}\leftarrow3s_{1/2}$ and $4s_{1/2}\leftarrow
3p_{1/2,3/2}$, respectively.

The effect of a quadratic phase modulation in frequency domain of
the form
\begin{equation}
\varphi(\omega) = \frac{\chirp}{2}(\omega-\omega_0)^2
\end{equation}
is described in time domain by a modulated linearly polarized laser
electric field $E(t)$ given as \cite{Wollenhaupt2007Springer}
\begin{equation}
E\left( t\right) =2Re\left\{E^{+}\left(t\right)\right\},
\end{equation}
where for the positive-frequency part we have \be E^{+}\left(
t\right) =\frac{E_{0}}{2\gamma^{1/4}}e^{-\frac{t^{2}}{4\beta
\gamma}}e^{i\omega_0 t} e^{i(a t^2-\varepsilon)} \ee with \bea
\varepsilon &=&\frac{1}{2}\arctan \frac{\chirp}{2\beta},\\
\beta &=&\frac{\Delta t^2}{8\ln 2},\\
\gamma &=&1+\left( \frac{\chirp}{2\beta}\right) ^{2},\\
a&=&\frac{\chirp}{8\beta ^{2}\gamma} \eea resulting in the
time-dependent instantaneous laser frequency \be \label{Eqn:Chirp}
\omega(t) = \omega_0 + 2 a t. \ee
Here $\Delta t$ denotes the FWHM of the intensity $I(t)$ of the
unmodulated pulse, $\omega_0$ is the laser carrier frequency and
$\chirp$ is the chirp parameter to be varied.

We define a reference Rabi-frequency $\Omega(t)=\Omega_0 f(t)$,
where $f(t)$ is the laser electric field envelope \be f\left(
t\right) =\frac{\exp\left(-\frac{t^{2}}{4\beta
\gamma}\right)}{\gamma ^{1/4}}. \ee

\ack Discussions with P. Lambropoulos and F. Faisal regarding the
measured photoelectron signals assigned to two-photon ionization
from state $3p$ are gratefully acknowledged. The Kassel group
furthermore acknowledges financial support by the Deutsche
Forschungsgemeinschaft DFG. This work has also been supported by
European Commission projects EMALI and FASTQUAST, and the Bulgarian
NSF grants VU-F-205/06, VU-I-301/07, D002-90/08, and IRC-CoSiM.

\end{document}